\documentclass[runningheads,a4paper]{llncs}
\usepackage{amssymb}
\usepackage{graphicx}
\usepackage{caption}
\usepackage{subcaption}
\usepackage{booktabs}
\usepackage[table,xcdraw]{xcolor}

\begin{document}
\mainmatter

\title{Improving tourist experience through an IoT application based on FatBeacons} 

\titlerunning{Improving tourist experience through an IoT application}

\author{Mois\'{e}s Lodeiro-Santiago\inst{}
\and Pino Caballero-Gil\inst{}
\and C\'{a}ndido Caballero-Gil\inst{}
\and F\'elix Herrera Priano\inst{}
}

\authorrunning{M. Lodeiro-Santiago, P. Caballero-Gil, C. Caballero-Gil,  F. Herrera Priano}

\institute{Departamento de Ingenier\'{i}a Inform\'atica y de Sistemas, \\Universidad de La Laguna, Tenerife. Spain\\
\email{\{mlodeirs, pcaballe, ccabgil, fpriano\}@ull.edu.es}
}

\toctitle{Lecture Notes in Computer Science}
\tocauthor{Authors' Instructions}
\maketitle
\keywords{Bluetooth Low Energy, FatBeacon, IoT, Tourism, Point of Interest}

\section*{Abstract}
This paper describes the use of a new extension of the Bluetooth connection protocol,  called  FatBeacon, which faces the problem of obtaining information where no Internet connection is  available. Rather than advertising a URL to load a web page,  the FatBeacon protocol has the ability to broadcast any basic web contents actually hosted on the device. In particular, FatBeacons are here used to improve the tourist experience in places with no Internet coverage through a new application of the Internet of Things (IoT). Thanks to the fact that the web content is emitted by the own FatBeacon, any  smartphone with Bluetooth Low Energy (BLE) can be used to receive touristic information, even in uncovered areas, such as rural or mountain destinations.  
This work does not only show the applicability of the new FatBeacon  protocol, but it also presents a performance comparison of different BLE technologies used for similar  touristic applications. 

\section{Introduction}
\begin{figure}[!htb]
    \centering
    \includegraphics[width=\textwidth]{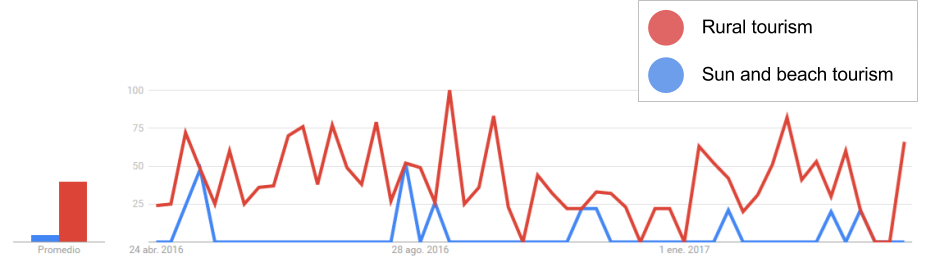}
    \caption{Comparison between searching terms}
    \label{fig:datosturismo}
\end{figure}

Tourism has become one of the major economic activities in most countries around the world. 
Tourism offers have changed a lot in recent years mainly due to the increase in the interest of tourists in new outdoor  activities. 
A proof of that is  searching terms like ``Rural tourism''  have increased notably  in contrast to the classic ''Sunshine and beach tourism''  \cite{turismo}  to the point of overcoming it (see Fig. \ref{fig:datosturismo}). 
Tourism has been also affected in recent years due to the widespread of many  new technologies. That is why, even in rural tourism, it is frequently common to combine it with activities that demand a certain use of technology. In particular, the potential of Internet of things (IoT) in tourims has been proven through many different applications and proposals.

Point Of Interest (POI) is the name given to places such as natural landmarks, historic monuments, museums and other places geolocated at a specific position  considered of a high interest for tourists. 
In many cases, tourists can not find basic information about the place, or about possible cultural or gastronomic activities that can be done there. In recent years the POIs have been technified, using for example Bluetooth technology in autonomous beacons.
As described in Section \ref{sec:beaconImplementation}, those Bluetooth beacons are continuously broadcasting a Uniform Resource Locator (URL) pointing to a web page. Android devices  that have a Bluetooth 4.0 version or higher will receive this signal with the contained URL and will automatically   proceed to download the content of this URL \cite{7066499}. 
The main disadvantage and problem of this kind of technology is that it requires an Internet connection in the  terminal. This requirement is a real problem in many cases  because although in cities it is common to have coverage 2, 3 and 4G, there are many remote places where no Internet connection is  available. 

This paper proposes an IoT-based application of the new protocol presented in the summer of 2016, known as FatBeacon. This new protocol allows the broadcast of web content (not to be confused with a  URL) that may contain HTML, CSS, JavaScript and other resources like images, vectors, etc. The main advantage of this new protocol is that the end user does not need to have an Internet connection to be able to see the content on its mobile terminal.

Bluetooth  is nowadays a well-known technology thanks to its many different applications. Its use has been enhanced through several  improvements in the corresponding protocols. Especially in recent years, with its integration into beacons, Bluetooth has been proposed as a solution to optimize resource allocation and indoor location \cite{alvarez2016optimizing} or in combination with technologies such as unmanned aerial vehicles (commonly known as drones) for the location of missing people \cite{Lodeiro-Santiago2016}. It has also been used in the smart-city concept \cite{kim2016smart}, which has been aroused in the last years, to broadcast information of POIs, products, shops, etc. Until today, all applications and projects have made use of Bluetooth technology for POIs as a sensor dependent on other technologies, such as the Internet,  to obtain information, as shown in \cite{shibata2016tourist}.
The new FatBeacon protocol has the advantage of allowing  connectivity in areas without Internet signal. However, it has  disadvantages such as  high power consumption and low battery life. However, these problems can be reduced or mitigated  using other power energy like solar cells.

Until now, there was no  application that allowed to obtain information from any POI without making use of an Internet connection   \cite{ruta2016physical}. 
This paper considers the solution to this problem using the new FatBeacon protocol applied for this purpose. In addition, it provides a comparison of different beacon protocols by measuring their performance.

The present paper is organised as follows.  Section \ref{sec:preliminaries}  describes the network coverage issue. 
A brief overview of  Bluetooth technology is presented in Section \ref{sec:beaconImplementation}. 
In Section \ref{sec:casestudy}, a case study and a detailed proof of concept are presented. 
Finally, in  Section \ref{sec:conclusiones}, the work is closed with some conclusions and future lines.

\section{Preliminaries}
\label{sec:preliminaries}
The use of mobile networks is becoming more common. In fact it is already considered almost vital in the daily life of a common user. This type of network is deployed by establishing telephone antennas located at strategic points to attempt to cover a maximized radio and mobile coverage area. At present, connection bands are divided into 2G, 3G, 4G and even in the near future (2018), the use of 5G connections will be in regular use.
The range of coverage of data networks is usually inversely proportional to their speed. For example, the 2G connection has low speed, compared to the other connections, but has a fairly wide coverage.
The majority of users who have an Internet connection on their mobile phones use them to obtain information about places, restaurants, social networks, etc.
However, in hiking tourism, this type of information, such as weather, track status, local flora and fauna, recommendations or activities related to trekking, is not available in many cases.
The FatBeacon protocol is used here to help solve this problem.

\begin{figure}[!htb]
    \centering

    \begin{subfigure}[b]{0.4955\textwidth}
        \includegraphics[width=\textwidth]{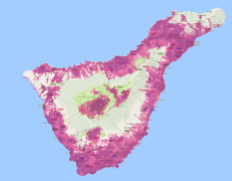}
        \caption{2G coverage map}
        \label{fig:sendero0}
    \end{subfigure}
    \begin{subfigure}[b]{0.4955\textwidth}
        \includegraphics[width=\textwidth]{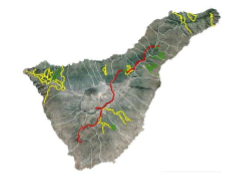}
        \caption{Homologated walking trails}
        \label{fig:sendero1}
    \end{subfigure}
    ~ 
    \begin{subfigure}[b]{0.4955\textwidth}
        \includegraphics[width=\textwidth]{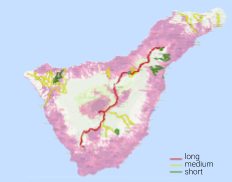}
        \caption{Walking trails over 2G coverage map}
        \label{fig:sendero2}
    \end{subfigure}
    \begin{subfigure}[b]{0.4955\textwidth}
        \includegraphics[width=\textwidth]{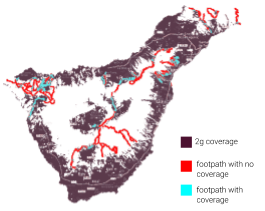}
        \caption{Walking trails with/without 2G coverage}
        \label{fig:sendero3}
    \end{subfigure}
    \caption{Example of trail routes and signal coverage}\label{fig:tenerife}
  \label{fig:senales_t}
\end{figure}

As an example,  we now consider the case of the Spanish community of Santa Cruz de Tenerife, where many outdoor  activities are offered for hikers \cite{diariodeavisos} in different varieties of trails approved and enabled for their use. Some of them are even enabled for people with reduced mobility. At present, the trails (see Fig.\ref{fig:sendero1}) are classified as follows:

\begin{itemize}
    \item Green: These are local trails, less than 10 km, approved and verified by the issuing authority.
    \item Yellow: These are short distance trails, with a length between 10 and 50 km, but generally with quite a few slopes.
    \item Red: These trails are long-haul, over 50 km, so they are designed to make them in several days.
\end{itemize}

Fig.\ref{fig:sendero0} shows that a large part of the area is not covered even by the wider range (2G). This implies that approximately 40\% of the island is inaccessible to Internet connections.  Fig. \ref{fig:sendero2}  shows an overlap of the trails layer over the 2G map layer (see Fig.\ref{fig:sendero0}). Working with this composition and using a technique of difference of layers of the image where if there is a point in the layer A that is present in the layer B, that point is painted   a certain colour, as seen in Fig.\ref{fig:sendero3},  it is possible to  graphically visualize the total  area of trails that have 2G coverage, and which of them do  not have it. In this case, we can see the red colour   indicating the number of trails (60\% of the trails section) that is not covered coverage. On the other hand,   40\% does have, at best, mobile data coverage.

The signs that a tourist  can find along the trail are those that can be seen in   Fig. 2 (among others). For example,  we can find logs with the colours of the trails or vertical signs that can give us a notion or guide of the site. These types of signs often deteriorate due to various climatic conditions or due to bandy acts.

\begin{figure}[!htb]
    \centering
    
    \begin{subfigure}[b]{0.45\textwidth}
        \includegraphics[width=\textwidth]{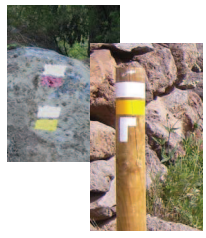}
        \caption{Painted on stakes to indicate trail}
    \end{subfigure}
    ~
    \begin{subfigure}[b]{0.45\textwidth}
        \includegraphics[width=\textwidth]{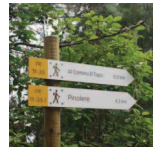}
        \caption{Vertical signals}
    \end{subfigure}
    \label{fig:senales_t}
    \caption{Guide signals}
\end{figure}

\section{FatBeacon-based implementation}
\label{sec:beaconImplementation}
BLE  devices that  operate on batteries and offer specific functionality are generally referred to as beacons.
 These devices emit a spherical BLE signal that, depending on the used protocol, can broadcast different contents, although in the majority of cases it emits a web address URL. 
Apart from the URL, there are other parameters that are emitted in each packet, such as the transmission power (TX power), which is a fixed emission value. This value determines the approximate emission range of each beacon and the battery consumption. The emission range has a direct relationship with the TX value so that a higher TX value implies that the signal will be emitted strongest and will reach a longer distance. 
However, this implies a greater consumption of battery, which with a lower level of TX would not happen. The emission frequency is the period between two consecutive transmissions. This can also affect battery consumption, since   a higher frequency means more battery consumption. Frequency times are generally expressed in milliseconds (ms), so that a low  value (higher emission frequency) could be 100ms (10 times of emission in one second) while a high frequency (500ms for example) would imply the emission of two transmissions per second. 
Mainly the beacon's life (in battery consumption) depends on these two values, so that the more it requires emission, the faster the battery will be spent. Generally, beacons are powered by lithium batteries such as CR2032 because they take up little space and have enough strength to power up the internal circuit. There are also models powered by a sunlight panel \cite{cyalkit} (among others) that also has a charge accumulator to be able to work at night with the energy obtained during the day. 

As explained, the BLE signal is emitted using a data transmission protocol. Currently, the best-known protocols are the protocols: iBeacon \cite{iBeacon}  developed by Apple \cite{Newman2014}, AltBeacon \cite{AltBeacon} designed by Radius Networks, and  Eddystone \cite{EddyStone} created by Google. The main difference betwee these three protocols and the new FatBeacon protocol (also created by Google) is that the latter can be used to broadcast much more content, making use of a connection by the Bluetooth from the transmitter to the receiver. 
This protocol is still in an experimental state. The detection operation is identical to the other presented protocols (such as Eddystone), except that, together with the packages with different parameters, an internal HTML content is emitted. The main advantage of this is that it makes the user being independent of an Internet connection (and related to it, independent of any antenna coverage).

The main disadvantage of emitting HTML content through this new protocol is that the stream emitted must be atomic (indivisible). Usually, when creating a web page the common thing is having the styles, JavaScript scripts, images and HTML code separately in order to have a cleaner code and to have the resources separately. In this case, to transmit HTML content using FatBeacon it is necessary that everything is inside the HTML. This is solved by including CSS and JavaScript scripts in the document's \verb|<head>...</head>|  header. In the case of images, icons and other resources that have to be included in the web, a previous base64 encoding \cite{base64} is necessary for their inclusion as a string next to the \verb|<img src="data:image/format;base64,/9j/2306AQS..."/>| image tag.

\begin{figure}[!htb]
    \centering
    \begin{subfigure}[b]{0.45\textwidth}
        \includegraphics[width=\textwidth]{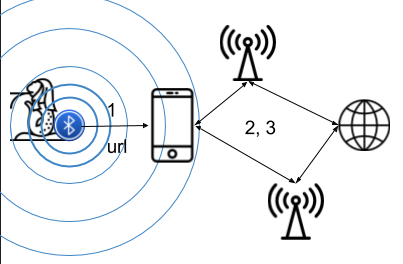}
        \caption{Actual use of BLE Beacons}
        \label{fig:eddy}
    \end{subfigure}
    ~
    \begin{subfigure}[b]{0.45\textwidth}
        \includegraphics[width=\textwidth]{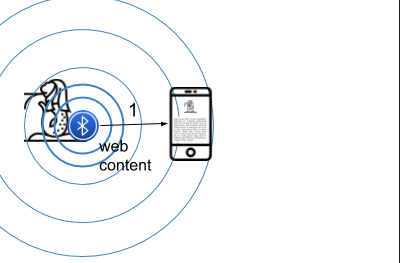}
        \caption{New use of BLE Beacons using FatBeacon  protocol}
        \label{fig:fat}
    \end{subfigure}
    
    \caption{Differences between  Eddystone  and   FatBeacon protocols}\label{fig:bles}
\end{figure}

In the following example (see Fig.\ref{fig:bles}) a diagram is shown divided into two parts (Fig.\ref{fig:eddy} and Fig.\ref{fig:fat}). This scheme differentiates the way that the beacons are normally used and how the new FatBeacon  protocol works (left and right respectively). The first image shows how a beacon   emits a web address (URL) that a smartphone receives through Bluetooth. Automatically, the smartphone, making use of an Internet connection, (mostly using 2G, 3G or 4G networks) downloads the content of the web and shows a notification on the screen. In the second image, where the FatBeacon protocol is used, steps 2 and 3 of Figure.\ref{fig:bles} do not exist. This is because, with the new FatBeacon protocol, web content (HTML content) is emitted directly by using a Bluetooth connection instead of downloading the contents from a web server.

In the following, some results are shown  of some comparative tests to compare the efficiency of the new protocol related to the others. Likewise, a comparison of load times for different connection types, different sizes and different distances is also made to determine a degree of correlation between the test and results.
The way to obtain the different speeds for BLE 4 has been  empirical, making use of two mobile devices, one as emitter and the other one as receiver (because there are currently no physical devices that implement the protocol FatBeacon). For this, as an emitter, a Samsung Galaxy S7 \cite{sgs7} (model SM-G930F) has been chosen. As receiver, a Xiaomi Mi 5s \cite{mi5s} smart-phone (model 3G/64GB) has been used. About the software, a custom version of the official Google Physical Web application (available to the public in \cite{physicalwebgh}) has been used. 

The changes made with respect to the original have been:

\begin{itemize}
    \item Modify the \verb|BluetoothSite.java| adding the following variables \verb|private| \verb|long start_time = 0;| and \verb|private long end_time = 0;|.
    \item In the \verb|connect| function was added \verb|start_time = System.nanoTime();|
    \item In the \verb|close| method, it was added \verb|end_time = System.nanotime();| and \verb|double difference = (end_time - start_time)/1e6;| at the end of the function (with a \verb|System.out.println(difference)| to check the result).
\end{itemize}

The file where the HTML content was stored was also modified to fit the following sizes (in kilobytes): 10, 20, 40, 100 and 200. Having this when running the application (having the receiver smart-phone connected to the emulation terminal) it is possible to see the time it took to transfer data using FatBeacon from one terminal to the another one. 

Other transfer speeds (BLE5\cite{ble5}, 2G and 3G \cite{3gpp}) have been extracted from the speed specifications of each type of connection and considering that \verb|1 byte =| \verb|8 bits|, what means that if the speed is 1 Mbit/s then, the real ratio is 0,125 MB/sec.

The resulting download times, taken by each kind of connection and sizes, are shown in the next Table.\ref{tab:speeds}.

    \begin{table}[!htb]
        \centering
        \caption{Download speeds for different protocols and sizes (in second/s)\\** Based on the BLE5 specifications\\*** Based on 3GPP specifications and without considering the TCP connection delay}
        \label{tab:speeds}
        \begin{tabular}{@{}lllll@{}}
        \toprule
              & BLE 4 & BLE 5* & 2G** & 3G** \\ \midrule
        10kb  & 5.21  & 1.30   & 5    & 0    \\
        20kb  & 8.82  & 2.20   & 11   & 0    \\
        40kb  & 7.43  & 1.85   & 23   & 0    \\
        100kb & 15.18 & 3.79   & 58   & 2    \\
        200kb & 28.14 & 7.03   & 107  & 4   
        \end{tabular}
    \end{table}

The connection times for BLE 4 have been extracted from Table.\ref{tab:tiempokb} performed in a related way (empirical) maintaining a distance between the transmitter and receiver of 1 meter. For each size of web page, five measurements were made. Then, discarding the best and the worst result, a mean of the remaining values was performed. When performing this experiment, a rare case was that the first result of each size was the worst, which could indicate that the RAM load of the web is done on the fly keeping it for the succeeding results.

\begin{table}[!htb]
\centering
\caption{Times at 1m}
\label{tab:tiempokb}
\begin{tabular}{@{}llllll@{}}
\toprule
1m & 10kb   & 20kb    & 40kb     & 100kb   & 200kb   \\ \midrule
1           & 4.0498 & 6.1178  & 4.3513   & 12.5816 & 15.8346 \\
2           & 4.4163 & 7.5264  & 7.2030   & 12.7050 & 23.4155 \\
3           & 5.2194 & 8.8279  & 7.4392   & 15.1869 & 28.1433 \\
4           & 6.8424 & 12.0863 & 8.6729   & 18.4735 & 33.7344 \\
5           & 7.3336 & 14.4279 & 10.35.77 & 74.8106 & 87.9002 \\ \midrule
Median      & 5.5219 & 8.8279  & 7.7492   & 15.1869 & 28.1433 \\ \bottomrule
\end{tabular}
\end{table}

According to the results shown in Table.\ref{tab:tiempokb}, the correlation index (\verb|0.9468|) between size and speed indicates that  a very strong dependence exists between size and speed of transmission.

In order to determine whether the transfer time depends on the distance, another empirical experiment was done, establishing a static web page size (40kb approximately) and a variable distance. The experiment was performed for each test, in a straight line without obstacles through and without factors that could attenuate the strength of the signal such as solar rays, wind, etc. resulting in the values   shown in  Table.\ref{tab:distance40} (where the results are ordered from best to worst for each distance).

\begin{table}[!htb]
\centering
\caption{Checking that the transfer time depends on the distance}
\label{tab:distance40}
\begin{tabular}{@{}lllll@{}}
\toprule
40kb              & 1m     & 5m     & 10m    & 15m    \\ \midrule
1 best result                                                                                 & 4.351  & 4.237  & 6.283  & 7.769  \\
2                                                                                             & 7.203  & 4.539  & 6.668  & 8.001  \\
3                                                                                             & 7.439  & 6.718  & 7.117  & 8.075  \\
4                                                                                             & 8.673  & 7.614  & 8.668  & 10.235 \\
5 worst result                                                                                & 10.358 & 10.246 & 30.187 & 10.784 \\ \midrule
\begin{tabular}[c]{@{}l@{}}Median\\ (discarding the best\\ and the worst result)\end{tabular} & 7.439  & 6.718  & 7.117  & 8.075  \\ \bottomrule
\end{tabular}
\end{table}

In the case shown in Table.\ref{tab:distance40}, the correlation index (\verb|0.6851|) indicates that exist a relationship between the distance and the speed.

\begin{figure}[!htb]
    \centering
    \begin{subfigure}[b]{0.45\textwidth}
        \includegraphics[width=\textwidth]{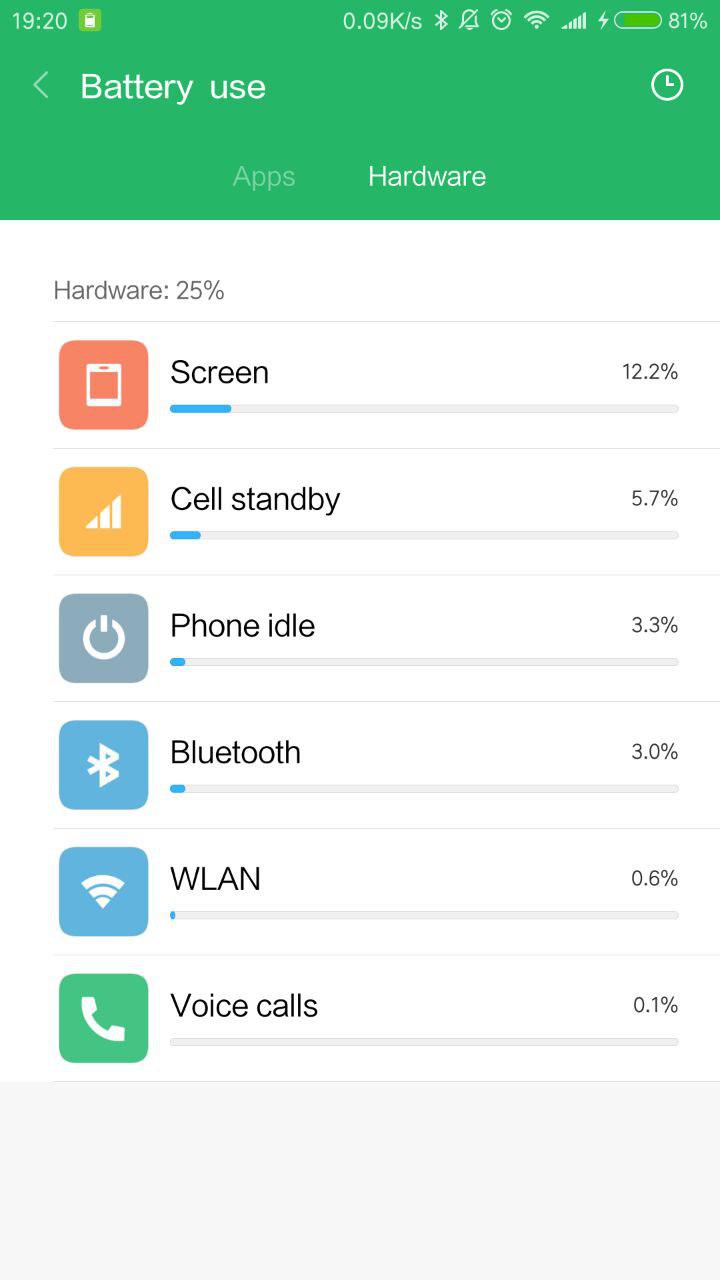}
        \caption{Battery use}
        \label{fig:tigessraa}
    \end{subfigure}
    ~
    \begin{subfigure}[b]{0.45\textwidth}
        \includegraphics[width=\textwidth]{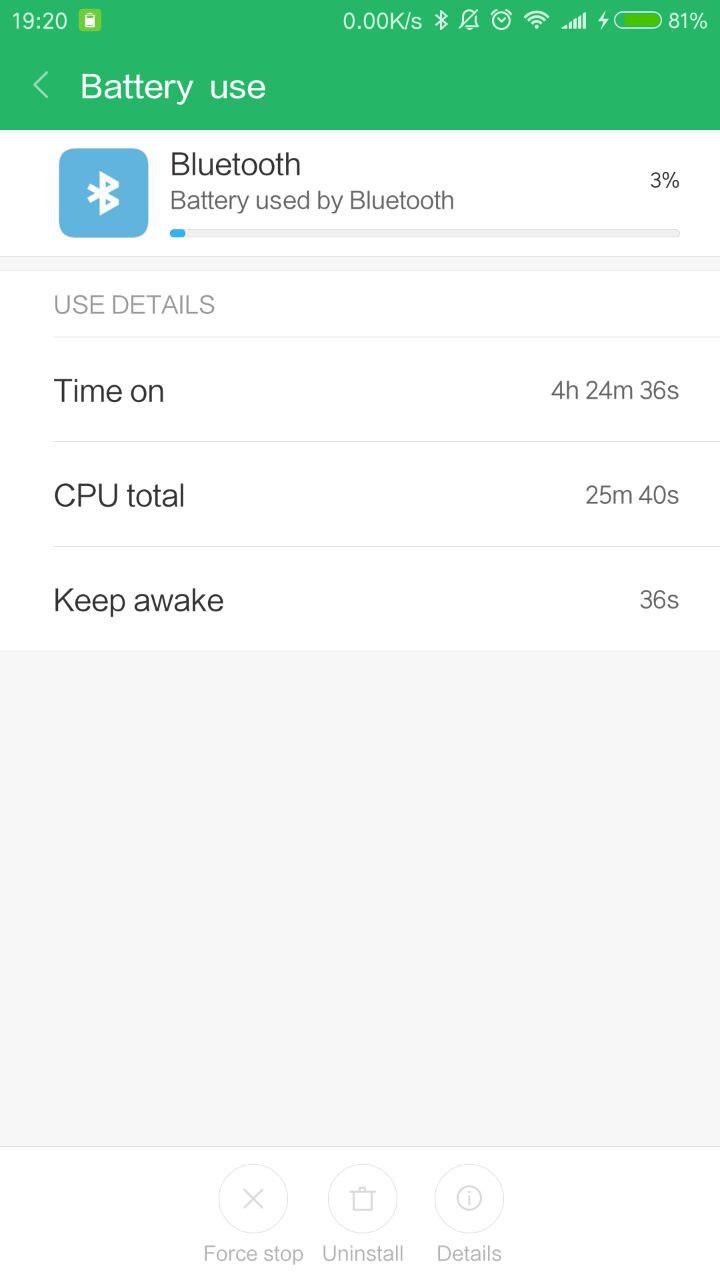}
        \caption{Bluetooth battery consumption}
        \label{fig:mousexssx}
    \end{subfigure}
    
    \caption{Bluetooth battery consumption}
    \label{fig:btbateria}
\end{figure}

Another great advantage of using a Bluetooth protocol in comparison with other connections, such as 2G, is that the battery consumption is quite lower, such as it is shown in the  screen-shots of  Fig.\ref{fig:btbateria}. Keep in mind that these data can change according to the user profile.

\section{Case Study}
\label{sec:casestudy}
\begin{figure}[!htb]
    \centering
    \begin{subfigure}[b]{0.55\textwidth}
        \includegraphics[width=\textwidth]{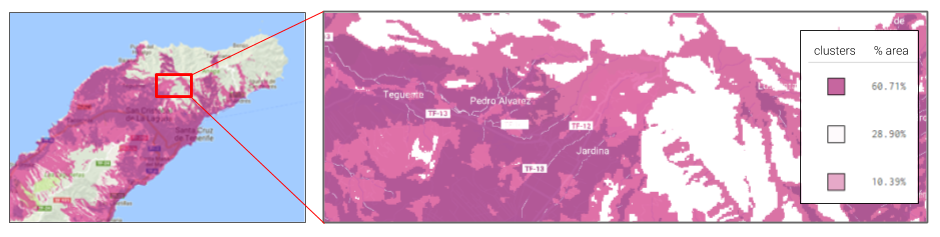}
        \caption{Trail location and 2G map coverage}
        \label{fig:area2g}
    \end{subfigure}
    ~
    \begin{subfigure}[b]{0.4\textwidth}
        \includegraphics[width=\textwidth]{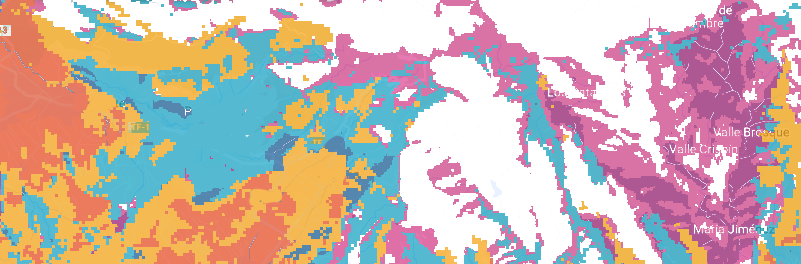}
        \caption{2G, 3G and 4G map coverage (overlayed)}
        \label{fig:area4g}
    \end{subfigure}
    \begin{subfigure}[b]{0.55\textwidth}
        \includegraphics[width=\textwidth]{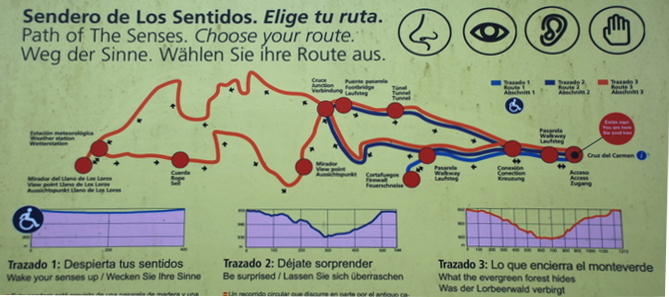}
        \caption{Entrance trail sign}
        \label{fig:senderocartel}
    \end{subfigure}
    \begin{subfigure}[b]{0.4\textwidth}
        \includegraphics[width=\textwidth]{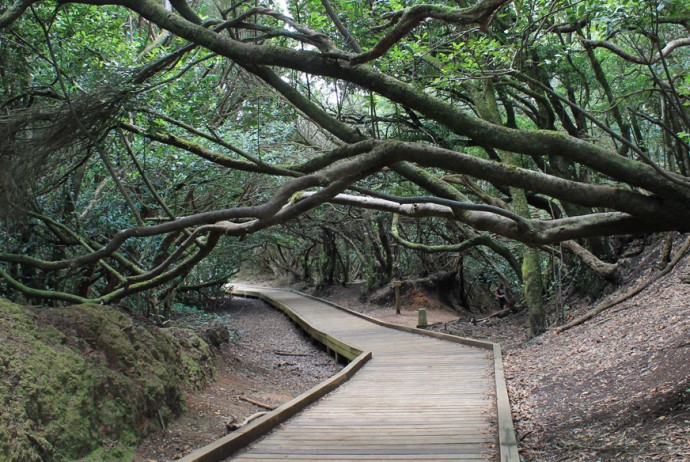}
        \caption{Picture of the grove}
        \label{fig:senderofoto}
    \end{subfigure}
    \caption{Trail example}

\end{figure}

As presented in Section \ref{sec:preliminaries}, one of the main problems with information on the fly is mobile coverage in many areas. In particular, in this case of use, the hypothesis of not having a connection at the entrance of a path covered by trees has been adopted. In normal situations in this trail, there could be a connection 2G as maximum (see Fig.\ref{fig:area2g} and Fig.\ref{fig:area4g}), if it is interposed any kind of vegetation or obstacles (trees, for example) the 2G connection turns into inexistent.

In this case, the practical tests were carried out on the ``Sendero de los Sentidos'' \cite{sentidos} in the island of Tenerife, in Spain.
This trail is very popular among visitors that come to Tenerife due to its low level of physical requirement, its accessibility (it is possible to arrive by car at the entry), access to people with reduced mobility, its diversity in flora and sensory experiences that are offered (see Fig.\ref{fig:senderocartel}). A feature of this trail is that its path is covered by trees (see Fig.\ref{fig:senderofoto}) so, any mobile coverage in the area is almost nonexistent.

\begin{figure}[!htb]
    \centering
    \begin{subfigure}[b]{0.475\textwidth}
        \includegraphics[width=\textwidth]{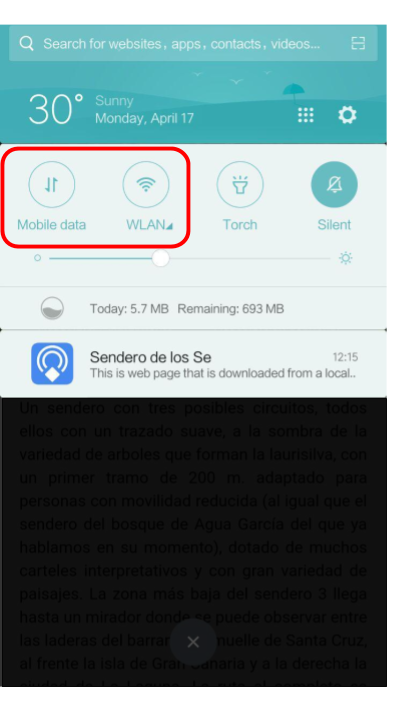}
        \caption{Content notification emitted by a  beacon}
        \label{fig:fb1}
    \end{subfigure}
    ~
    \begin{subfigure}[b]{0.475\textwidth}
        \includegraphics[width=\textwidth]{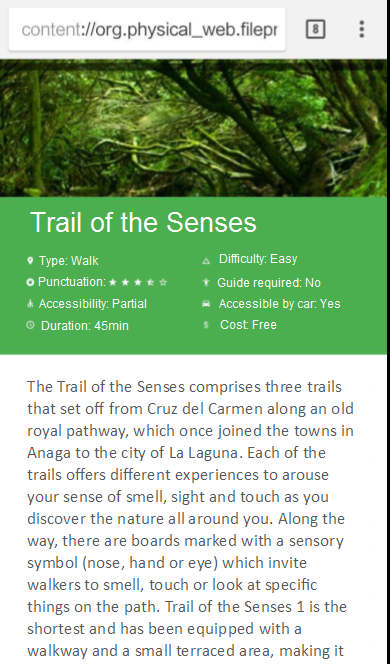}
        \caption{Web content emitted by a  beacon}
        \label{fig:fb2}
    \end{subfigure}
    \caption{BLE FatBeacon emission}

\end{figure}

After validating that the coverage was non-existent in the area, it was decided to disable connections referring to Internet data and Wi-Fi (to save battery). At the beginning of the trail, the emitter was placed emitting a 40kb (approximately) web content. When a user approaches a reasonable distance, the receiver picks up the signal emitted by the transmitter showing a notification in the panel of the smart-phone (see Fig.\ref{fig:fb1}). When the user clicks on the notification, a message of loading content appears to later open the web browser with the content emitted by the sender (see Fig.\ref{fig:fb2}).

\section{Conclusions}
\label{sec:conclusiones}
In this paper, the new protocol (still in pre-release phases) FatBeacon has been presented as an alternative that does not require any data connection to download a beacon web information.   In addition, a comparison has been made between said technology and other known BLE protocols, such as  Eddystone, iBeacon or AltBeacon. 
This new protocol offers the possibility of sending static HTML content through a Bluetooth connection so that people do not need to have data connectivity on their smartphones or an Internet connection to receive information from isolated POIs. 
This work shows a study of the efficiency of battery consumption of a device with Bluetooth Low Energy compared to others (like WiFi and data connection). In addition to this, empirical data of the efficiency of the new FatBeacon protocol is shown for differentfactors, such as distances, web page sizes, etc.  and has been compared with other BLE protocols such as BLE5. 
In this comparison, other forms of data transfer (having Internet) have been contrasted, considering different speeds like 2G, 3G and higher versions of Bluetooth as the integrated 5 in its low energy version.
The main conclusion of the experiments presented in this paper is the great utility of this new protocol and the advantages that it offers in front of its predecessors due to its versatility, low consumption and low cost of deployment. 
As future work,  a wider comparison among BLE protocols will be made. 
Furthermore, the idea of creating a physical device based on components such as Arduino powered by a solar panel, which is  able to broadcast  by FatBeacon web content to offer tourist information in smart cities and isolated environments.

%
%
\section*{Acknowledgments}
Research supported by the Spanish Ministry of Economy and Competitiveness, 
the  FEDER Fund, and the CajaCanarias Foundation, under Projects 
TEC2014-54110-R, RTC-2014-1648-8, MTM2015-69138-REDT and DIG02-INSITU

%
%
\bibliography{bibliography}
\bibliographystyle{ieeetr}

\end{document}